\documentclass[%
 reprint,
superscriptaddress,
 amsmath,amssymb,
 aps,
]{revtex4-2}

\usepackage{graphicx}
\usepackage{dcolumn}
\usepackage{bm}

\usepackage{xcolor}

\begin{document}

\preprint{APS/123-QED}

\title{Three-dimensional theory of superradiant free-electron lasers}

\author{River~R.~Robles}
\email{riverr@stanford.edu}
\affiliation{SLAC National Accelerator Laboratory, Menlo Park, CA 94025}
\affiliation{Stanford University, Department of Applied Physics, Stanford, CA 94305}
\affiliation{Stanford PULSE Institute, SLAC National Accelerator Laboratory, Menlo Park, CA 94025, USA}

\author{Luca Giannessi}
\affiliation{Elettra-Sincrotrone Trieste, 34149 Basovizza, Trieste, Italy}
\affiliation{ENEA C.R. Frascati, 00044 Frascati (Roma), Italy}

\author{Agostino Marinelli}
\email{marinelli@slac.stanford.edu}
\affiliation{SLAC National Accelerator Laboratory, Menlo Park, CA 94025}
\affiliation{Stanford PULSE Institute, SLAC National Accelerator Laboratory, Menlo Park, CA 94025, USA}
\date{\today}

\begin{abstract}
The soliton-like superradiant regime of free-electron lasers (FEL) offers a promising path towards ultrashort pulses, beyond the natural limit dictated by the bandwidth of the high-gain FEL instability.
In this work we present a three-dimensional theory of the superradiant regime, including the effects of beam emittance and energy spread. Our work takes advantage of recent developments in non-linear FEL theory to provide a fully analytical description of  soliton-like superradiance.
Our theory proves the existence of a diffraction-dominated steady-state regime in which the superradiant peak power grows indefinitely while leaving the pulse duration and on-axis intensity almost unchanged. These results are in excellent agreement with three-dimensional simulations and are supported by recent experimental results at the Linac Coherent Light Source. This work advances non-linear FEL theory and provides a theoretical framework for the next generation of attosecond x-ray FELs.
\end{abstract}

\maketitle





\section{Introduction}

X-ray free-electron lasers (XFELs) are unrivaled in their ability to produce short, high power x-ray pulses \cite{emma2010first, ishikawa2012compact, kang2017hard, decking2020mhz, prat2020compact}. 
The recent development of attosecond XFELs \cite{duris2020tunable, zhang2020experimental, huang2017generating, marinelli2017experimental, prat2020compact} in particular has enabled the generation of isolated pulses with hundreds of attosecond duration and intensities surpassing table-top harmonic sources by six orders of magnitude. This enables novel applications combining attosecond time resolution with nonlinear spectroscopy \cite{o2020electronic} and single-particle imaging \cite{kuschel2022enhanced}. 
Achieving increasingly higher time-resolution and intensity requires a deeper exploration of non-linear FEL dynamics and the fundamental limits of short pulse generation.

In a high-gain FEL, a beam of relativistic electrons interacts with a co-propagating radiation pulse in a magnetic undulator \cite{pellegrini2016physics, huang2007review}.
A resonant interaction ensues, giving rise to a collective instability that amplifies exponentially the radiation power, and introduces a density modulation in the beam known as microbunching.
In this context, the amplification bandwidth of the XFEL is determined by the relative slippage of the electrons and the radiation. This poses a lower bound for the pulse duration on the order of the cooperation length, defined as the slippage length in one exponential gain-length \cite{bonifacio1994spectrum} (on the order of few fs to few hundred attoseconds at x-ray energies for typical experimental parameters).
Similarly, the power of a high-gain FEL is limited to its saturation value, which is achieved as the electron beam approaches full microbunching \cite{bonifacio1984collective}.
Both effects are described in the one-dimensional (1D) theory of FELs by the Pierce parameter $\rho$, which defines the bandwidth, efficiency, and gain-length of the system \cite{bonifacio1984collective}.
The cooperation length can be reduced and the saturation power increased by increasing the electron beam quality, but their ultimate values are always tied to the universal scaling of the FEL dynamics with the Pierce parameter.

One regime of the FEL bypasses these limitations: the superradiant mode \cite{bonifacio1985superradiant,bonifacio1989superradiance,bonifacio1991superradiant,giannessi2005nonlinear,yang2020postsaturation}. In this highly nonlinear regime, a short radiation pulse slips over the electrons at the speed of light, continuously extracting energy from fresh slices of the beam. In this regime, the interaction is so strong that the electrons become fully bunched as they interact with the rising edge of the pulse. This leads to amplification of the head of the pulse and absorption of the tail. The result is a soliton-like pulse that maintains a characteristic self-similar shape while growing in intensity and becoming shorter in the time domain.
Because the interaction is continuously sustained by new electrons, the superradiant mode has no saturation. 
Therefore, the peak power can grow indefinitely and the pulse duration can shrink well below the high-gain cooperation length.
Superradiance was first experimentally observed at infrared wavelengths \cite{watanabe2007experimental}, and was extended to the XUV range \cite{mirian2021generation} employing a harmonic cascade scheme \cite{giannessi2012high,giannessi2013superradiant}. More recently, superradiance has been observed in a cascaded x-ray FEL \cite{franz2024}, resulting in terawatt-scale attosecond pulses. These experimental results call for a deeper understanding of the superradiant mode and its ultimate limits.

According to the existing 1D theory of superradiant FELs \cite{bonifacio1990superradiant}, the peak power grows quadratically with the propagation distance $z$, while the pulse duration shrinks as $z^{-1/2}$. 
The 1D theory neglects diffraction, which has only been studied using numerical simulations \cite{giannessi:fel05-mooc001}. 
Since the on-axis intensity of a strongly diffracting field in free space decreases as $z^{-2}$,  the superradiant intensity gain on axis can be canceled almost exactly by diffraction losses. Therefore, we can expect diffraction to dramatically change the behavior of superradiant FELs if the interaction length exceeds the Rayleigh length of the radiation pulse.
In this work, we report a new diffraction-dominated superradiant regime of the free-electron laser. In this regime, the pulse, closely coupled to the electron beam microbunching, develops into a soliton-like mode qualitatively similar to the 1D soliton. However, the inclusion of diffraction causes the on-axis intensity and pulse duration to change more slowly than in the 1D limit as diffraction losses nearly balance on-axis amplification, and the growth of the peak power is ultimately linear with $z$. We study this regime first using 3D FEL simulations. We then develop a self-consistent, asymptotic theory that accurately describes the key properties of the superradiant pulse. Moreover, we find that the output pulse properties are weakly dependent on the initial conditions that launch the mode, suggesting superradiance as a route to short, intense, and stable pulses. Finally, we study semi-analytically the impact of beam emittance and energy spread.

\begin{figure*}[htb!]
    \centering
    \includegraphics{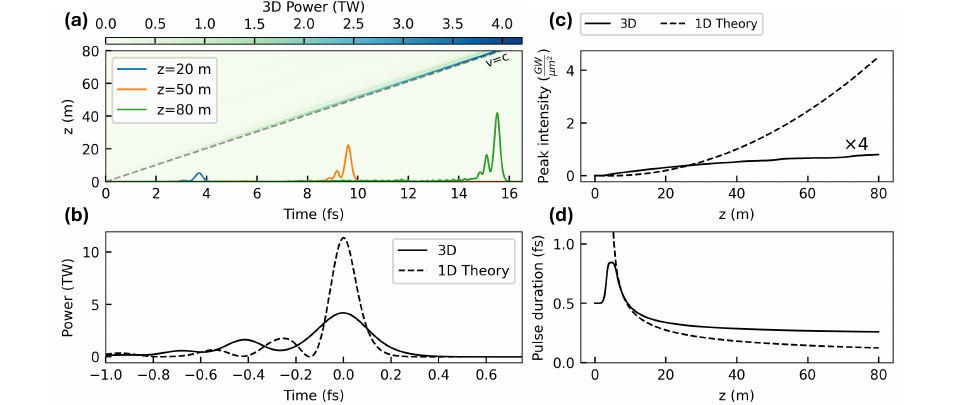}
    \caption{Basic differences between superradiance with (3D) and without (1D) diffraction. (a) shows the evolution of pulse power over a 80 meter undulator with lineouts of the temporal profile at three positions. (b) shows the final power profile for the 1D and 3D cases. (c) and (d) show the peak intensity and pulse duration versus undulator length.}
    \label{fig:1D_3D_comparison}
\end{figure*}

To introduce the qualitative features of the problem, Fig.~\ref{fig:1D_3D_comparison} shows the results from a 3D simulation using the GENESIS 1.3 v4 code \cite{reiche1999genesis,genesis4github} and the self-similar 1D solution starting from 0.5\% initial bunching (see, for example, \cite{kim2017synchrotron,piovella1991hyperbolic}). The parameters used in the simulations in all figures in the paper are reported in Table~\ref{tab:sim_parameters}. To start, the simulations consider zero emittance and energy spread in order to isolate the impact of diffraction from other complicating 3D and thermal effects. Furthermore, in all of the simulations, the beam particles are loaded without shot noise (``quiet-loaded") so as to highlight the superradiant physics in the asymptotic regime.

\begin{table}[h!]
    \centering
    \begin{tabular}{c|c}
        \hline
        Beam parameter & Value  \\
        \hline
        Current $I$ & 4 (2) kA\\
        Beam size $\sigma_r$ & 20 $\mu$m\\
        Energy $\gamma_0 mc^2$ & 5 GeV \\
        \hline
        Undulator parameter & Value\\
        \hline
        Period $\lambda_u=2\pi/k_u$& 3.9 cm \\
        Length $L_u$ & 80 (60) m\\
        Resonant photon energy $(\hbar k_rc)$ & 540 eV\\
        \hline
        Seed field parameter & Value \\
        \hline 
        FWHM power duration & 0.5 fs\\
        Peak intensity & 1.27 (6.37) MW/$\mu$m$^2$\\
        Spot size $w_0$ \cite{siegman1986lasers} & 70.7 $\mu$m
    \end{tabular}
    \caption{Parameters for simulations. Values given in parentheses are relevant for Fig.~\ref{fig:vs_beamsize}, \ref{fig:vs_z}, and \ref{fig:espread_emittance_scan} where applicable. Other differences in later simulations are explicitly stated in the text.}
    \label{tab:sim_parameters}
\end{table}

Panel (a) shows the evolution of the power profile in the 3D simulation with lineouts at three positions in the undulator. As the pulse propagates, the coherent gaussian seed gives rise to the expected superradiant profile composed of a primary peak followed by a sequence of weaker trailing peaks. Panel (b) shows the output power profile compared to the 1D self-similar superradiant solution. Panels (c) and (d) show the evolution of the on-axis intensity and full-width at half-maximum pulse duration in the two cases. The dashed curves showing the 1D results scale as $z^2$ and $z^{-1/2}$, respectively. In contrast, both the intensity and the duration exhibit a plateau in the 3D simulations. The result indicates that the pulse duration is not well described by the $1/\sqrt{z}$ dependence expected from the 1D theory. 
We note that in linear FEL theory, the effect of diffraction can be included in a pseudo 1D model by re-scaling the FEL equations to account for the longer gain length  \cite{giannessi:fel06-mopph026,xie1995design}. This approach does not work for the superradiant regime because of the lack of gain-guiding, as discussed in Appendix~\ref{app:scaled_rho} and illustrated in Fig.~\ref{fig:1D_vs_3D_with_correction}. Therefore a new theoretical approach is required to fully describe the physics at play.



\section{Theoretical results}

In what follows we derive an asymptotic model of superradiant FELs in the diffraction dominated regime. More complete details of the mathematical derivation are given in Appendix~\ref{app:derivations}. We expect this regime to be realized when the interaction length $z$ after the initial saturation is much larger than the typical length-scale of diffraction: $z \gg k_r \sigma_r^2$, where $k_r$ is the radiation wavenumber and $\sigma_r$ is the transverse beam size.
The starting point for our theoretical model is the 3D FEL wave equation (see, for example, \cite{huang2007review,pellegrini2016physics,kim2017synchrotron}):
\begin{align}
    \left(\frac{\partial}{\partial z}+\frac{1}{c}\frac{\partial}{\partial t}-\frac{1}{2ik_r}\nabla_\perp^2\right)E_x(z,t,\mathbf{r})=\kappa b(z,t)u(\mathbf{r}).
\end{align}
The coordinates $z$, $t$, and $\mathbf{r}$ refer to the distance along the undulator, time, and transverse coordinates, respectively. $E_x(z,t,\mathbf{r})$ is the slowly-varying envelope of the electric field, $b(z,t)=\langle{e^{i\theta_j}\rangle}$ is the bunching at a fixed location in the beam, where $\theta_j=(k_r+k_u)\bar{z}_j(t)-\omega_rt$ is the ponderomotive phase of the j-th electron, and $u(\mathbf{r})$ is the transverse profile of the beam, where $\int d\mathbf{r}u(\mathbf{r})=1$. Finally, $k_u$ is the undulator wavenumber, and the coupling term $\kappa=IK\text{[JJ]}/4\epsilon_0c\gamma_0$, where $K$ is the undulator strength parameter, $I$ is the beam current, $\gamma_0$ is the beam Lorentz factor, and $\text{[JJ]}$ is the Bessel coupling factor \cite{kim2017synchrotron,pellegrini2016physics,huang2007review}.
Note that we are assuming that the bunching factor does not have an explicit transverse dependence. This is appropriate in a diffraction-dominated case because the radiation mode is larger than the electron beam and the field generating the bunching is roughly uniform over the transverse extent of the electron distribution.

The physical picture presented by the simulations and by the intuitive arguments above suggests that in the diffraction-dominated superradiant regime the pulse propagates at the speed of light with nearly constant on-axis intensity. This is similar to the 1D regime, where the pulse and beam microbunching both propagate at a velocity asymptotically approaching the speed of light, but with growing on-axis intensity \cite{yang2020postsaturation}.
We therefore assume that the electron dynamics depend strongly on the variable $\zeta=z-ct$, and have only a weak explicit dependence on $z$. We change variables from $(z,t)$ to $(z,\zeta)$ and assume $b(z,t)\simeq b(\zeta)$. 
The resulting wave equation is formally equivalent to that of undulator radiation from a pre-bunched beam \cite{saldin2005simple} and has the following solution:
\begin{align}\label{eqn:field_in_terms_of_bunching}
    E_x(z,\zeta,\mathbf{r})=2ib(\zeta)k_r\kappa\int\frac{d^2\mathbf{k}}{4\pi^2} e^{i\mathbf{k}\cdot\mathbf{r}}\frac{\tilde{u}(\mathbf{k})}{k^2}\left(1-e^{\frac{ik^2z}{2k_r}}\right)
\end{align}
where we have introduced the Fourier transform of the bunch shape $\tilde{u}(\mathbf{k})=\int d\mathbf{r}e^{-i\mathbf{k}\cdot\mathbf{r}}u(\mathbf{r})$. The bunching $b(\zeta)$ can be determined self-consistently given this field and the equations of motion for the electrons. Written in terms of the relative energy deviation $\eta_j=(\gamma_j-\gamma_0)/\gamma_0$ and the ponderomotive phase $\theta_j$, the particles obey the pendulum equations:
\begin{align}
    \frac{d\theta_j}{dz}&=2k_u\eta_j\\
    \frac{d\eta_j}{dz}&=-\chi\left(E_xe^{-i\theta_j}+\text{c.c.}\right)
\end{align}
where $\chi=eK\text{[JJ]}/2\gamma_0^2mc^2$ and the field must be evaluated at the position of the $j$-th particle. In the diffraction-dominated regime, we assume that the field varies little over the beam area, and the particles move under the influence of the on-axis field.
Using Eq.~\eqref{eqn:field_in_terms_of_bunching}, for a gaussian beam with $u(\mathbf{r})=\frac{1}{2\pi\sigma_r^2}e^{-\frac{r^2}{2\sigma_r^2}}$, the on-axis field is given by:
\begin{align}
    E_x(z,\zeta,\mathbf{0})= ib(\zeta)k_r\kappa \frac{1}{2\pi}\ln\left(1-\frac{iz}{k_r\sigma_r^2}\right)
\end{align}
Because the on-axis field only depends logarithmically on $z$, we will neglect its $z$ dependence during the interaction of a given electron with the short radiation pulse. In that case, a set of self-consistent equations can be written in terms of the collective variables $b=\langle e^{i\theta_j}\rangle$ and $P=\langle \eta_je^{i\theta_j}\rangle$. These equations include non-linear terms which are critical for capturing the superradiant physics properly. They can be evaluated using the methods described in \cite{hemsing2020simple} as shown in Appendix~\ref{app:derivations}, with the end result being a single nonlinear differential equation for the bunching:
\begin{align}\label{eqn:duffing}
    \frac{d^2b}{dx^2}-b+3b|b|^2=0
\end{align}
where $x=\sqrt{\frac{k_r}{k_u}\frac{\chi\kappa}{2}\frac{|\ln(1-iz/k_r\sigma_r^2)|}{2\pi}}2k_r\zeta$. This equation is expected to be valid in the leading peak of the superradiant pulse. Equation~\eqref{eqn:duffing} has an approximate solution in terms of a hyperbolic secant \cite{hemsing2020simple}: 
\begin{align}
    b(\zeta)\simeq \sqrt{\frac{2}{3}}\text{sech}(x-x_0)\label{eqn:sech_bunching}
\end{align}
This is reminiscent of both the sech solution found for high-gain FELs in early saturation \cite{hemsing2020simple}, and in 1D superradiance \cite{piovella1991hyperbolic}. In what follows we will neglect the offset $x_0$, since it only shifts the temporal profile. With this we can write a nearly self-consistent expression for the radiation envelope,
\begin{align}
    E_x(z,\zeta,\mathbf{r})\simeq \text{sech}\left(\sqrt{\frac{k_r}{k_u}\frac{\chi\kappa}{2}\frac{|\ln(1-iz/k_r\sigma_r^2)|}{2\pi}}2k_r\zeta\right)\nonumber\\
    \times 2i\kappa k_r\sqrt{\frac{2}{3}}\int\frac{d^2\mathbf{k}}{4\pi^2}e^{i\mathbf{k}\cdot\mathbf{r}}\frac{\tilde{u}(\mathbf{k})}{k^2}\left(1-e^{\frac{ik^2z}{2k_r}}\right)\label{eqn:field_0th}
\end{align}
This expression allows us to make predictions about key parameters of the radiation pulse.  We will specifically highlight the peak power, the maximal on-axis intensity, and the pulse duration. We will derive explicit formulas for a Gaussian transverse electron beam profile $u(r)=\frac{1}{2\pi\sigma_r^2}e^{-\frac{r^2}{2\sigma_r^2}}$. More details and derivations for different beam distributions are found in Appendix~\ref{app:derivations}. First, the peak on-axis intensity and the full-width at half-maximum duration of the power and intensity asymptotically approach
\begin{align}
    I_\text{max} &= \frac{1}{3\epsilon_0c}\left[\frac{IK\text{[JJ]}k_r}{4\pi\gamma_0}\left|\ln\left(1-\frac{iz}{k_r\sigma_r^2}\right)\right|\right]^2\label{eqn:Imax_simple}\\
    \text{FWHM}_0 &= \frac{2\text{cosh}^{-1}(\sqrt{2})}{\omega_r}\sqrt{\frac{\gamma_0}{|\ln(1-iz/k_r\sigma_r^2)|}\frac{I_A}{I}\frac{1+\frac{K^2}{2}}{K^2\text{[JJ]}^2}}\label{eqn:fwhm_0th}
\end{align}
where $I_A\simeq17054$ A is the Alfvén current. The peak power asymptotically approaches
\begin{align}
    P_\text{max}&\simeq\frac{4\pi\sigma_r^2}{3\epsilon_0c}\left(\frac{IK\text{[JJ]}k_r}{4\pi\gamma_0}\right)^2\left[\frac{z}{k_r\sigma_r^2}\frac{\pi}{2}-2\ln\left(\frac{z}{k_r\sigma_r^2}\frac{e}{2}\right)\right]\label{eqn:Pmax_simple}
\end{align}
These equations are similar to those found for the diffraction-dominated post-saturation regime of the high-gain FEL \cite{fawley1996optical,saldin2005simple,schneidmiller2015optimization}, reflecting the physical mechanism of emission from a pre-bunched beam that is at the core of both regimes.

These formulas and equation~\eqref{eqn:field_0th} can be considered a 0th order solution since we neglected the $z$ dependence of the bunching, while the resulting field changes slowly with $z$. While solutions are already quite accurate, they can be further improved by iterating once more. To this end we insert the bunching equation~\eqref{eqn:sech_bunching} back into the original wave equation with its explicit $z$ dependence. This yields the following expression for the radiation field:
\begin{align}\label{eqn:field_with_zdep}
    E_x(z,\zeta,\mathbf{r})=\kappa\int\frac{d^2\mathbf{k}}{4\pi^2} e^{i\mathbf{k}\cdot\mathbf{r}}\tilde{u}(\mathbf{k})\int_0^zdz'b(z',\zeta)e^{-\frac{ik^2}{2k_r}(z'-z)}
\end{align}
This change leaves the peak power and intensity unchanged, but leads to the power and intensity profiles having different durations. This is because the radiation emitted on-axis is slowly compressing, whereas the off-axis radiation (which was emitted earlier in the amplification process) is fixed at its original duration. The correct FWHM is found by evaluating the integrals in Eq.~\eqref{eqn:field_with_zdep} as a function of $\zeta$. Numerical fitting formulas for the FWHM corrections can be written with the following fitting functions, which are accurate within 1\% over four orders of magnitude of the dimensionless diffraction parameter $z/k_r\sigma_r^2$ ($3<\frac{z}{k_r\sigma_r^2}<10^4$):
\begin{align}\label{eqn:fwhm_corrections}
    \frac{\text{FWHM}_\text{power}}{\text{FWHM}_0}&\simeq0.995+\frac{0.628}{0.27+|\ln(1-iz/k_r\sigma_r^2)|}\\
    \frac{\text{FWHM}_\text{intensity}}{\text{FWHM}_0}&\simeq0.973+\frac{0.254}{-0.64+|\ln(1-iz/k_r\sigma_r^2)|}
\end{align}

We have derived the expressions reported thus far in an asymptotic limit, using only a particular solution of the wave equation that vanishes at $z=0$, and neglecting the initial conditions that launch the superradiant mode. In reality, it takes a finite amount of time to launch the soliton mode starting from an initial seed. As a result, the field has an effective source point at $z = z_0 \neq 0$ (an analogous ambiguity exists in the 1D theory). Physically, $z_0$ is comparable to the length it takes for the seed pulse to fully microbunch a slice of the electron beam, and it is the point at which nonlinear superradiant dynamics become dominant. Evaluating the value of $z_0$ from an initial seed is beyond the scope of this paper, and we will use the results of simulations to estimate it for the examples reported here. In the simulation results that follow, we will show  our basic theory curves with $z_0=0$ as well as revised curves in which we substitute $z$ with $z-z_0$ and estimate the effective $z_0$ by a linear fit to the power from the simulations. Figure~\ref{fig:z0_vs_beamsize} in Appendix~\ref{app:z0_vs_beamsize} shows the dependence of $z_0$ on the beam size for the simulations to be shown in Figure~\ref{fig:vs_beamsize}. We find that, in practice, the formulas with $z_0=0$ are more than sufficient to estimate the pulse properties, but accounting for $z_0$ makes the agreement nearly perfect. 

\section{Simulation benchmarks}

\begin{figure}[h!]
    \centering
    \includegraphics{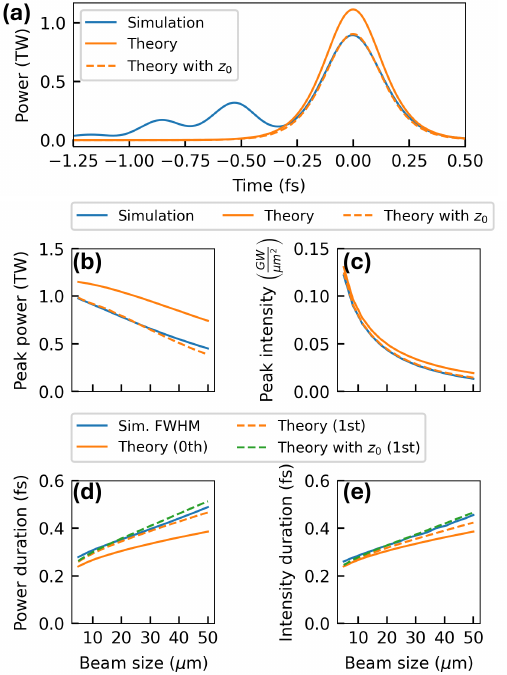}
    \caption{Comparison of theory with GENESIS simulations with various beam sizes. (a) shows the temporal power profile for an 11 micron beam size alongside the theoretical solutions. (b-d) show the output power, intensity, power duration, and intensity duration versus beam size.}
    \label{fig:vs_beamsize}
\end{figure}

We compare our results against 3D GENESIS simulations in Fig.~\ref{fig:vs_beamsize}. Panel (a) shows the power profile after a sixty meter long undulator with an 11 micron beam size. The solid orange curve is derived from the solution described by equations~\eqref{eqn:Pmax_simple}, including first-order corrections from equation~\eqref{eqn:field_with_zdep} for the pulse duration. The dashed orange curves are corrected using the $z_0$ estimate from the simulation as described above. 
Panels (b-e) show the peak power, peak intensity, and FWHM durations of power and on-axis intensity after a sixty meter undulator as a function of beam size. In the FWHM plots we have three theory curves: (0th) refers to the formulas not accounting for the slow $z$-dependence of the bunching equations~\eqref{eqn:fwhm_0th}, (1st) refers to revised formulas~\eqref{eqn:field_with_zdep}. We find excellent agreement in all parameters, in particular using the 1st order corrections for the FWHMs. Of particular note is that the intensity and duration predictions are quite good even without accounting for $z_0$. This is because of their logarithmic dependence on $z$, whereas the peak power which varies asymptotically linearly with $z$ experiences a larger shift when accounting for $z_0$.

\begin{figure}[h!]
    \centering
    \includegraphics{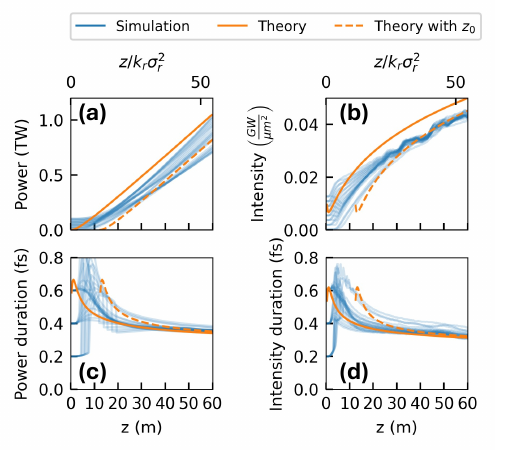}
    \caption{Comparison of theory with GENESIS simulations showing peak power, intensity, and pulse duration along the undulator for a variety of seed powers and durations, as compared to the asymptotic theoretical behavior. }
    \label{fig:vs_z}
\end{figure}

To explore the role of the initial seed field conditions on the ultimate behavior of the superradiant pulse, Fig.~\ref{fig:vs_z} shows simulation results for a $20$ $\mu$m beam size, and for a seed pulse with peak power ranging from 1 GW to 100 GW and FWHM duration between 0.2 and 0.6 fs. The same quantities are plotted as in Fig.~\ref{fig:vs_beamsize}, in the duration plots we now show only the 1st order theory curves. The $z_0$ used is the average for the 32 simulations. The on-axis intensity and FWHM durations of both power and intensity all asymptotically approach the same logarithmically varying curves, nearly independent of the initial conditions. The power varies more due to its linear dependence on $z_0$, a 12\% rms fluctuation in these simulations, which is still quite small considering the two order of magnitude variation in the seed power. This provides a path towards stabilizing the output of isolated attosecond XFELs, which suffer from significant shot-to-shot intensity fluctuations \cite{duris2020tunable}.

\section{Corrections for emittance and energy spread}

Now that the behavior of the 3D superradiant mode is well-understood for an ideal beam, we move to a discussion of the impact of non-ideal effects like energy spread and emittance. We consider a beam matched to a strong focusing lattice in the short cell approximation, in which the beam size is constant \cite{kim2017synchrotron}. In the theory, energy spread can be included simply by solving the pendulum equations for the bunching $b(x)$ allowing the energy coordinates $\eta_j$ to have a Gaussian energy distribution with a spread $\sigma_\eta$. Accounting for emittance, on the other hand, requires adding a term to the equation of motion for the ponderomotive phase $\theta_j$ which accounts for additional slowing of a particle's longitudinal velocity due to its thermal transverse velocity \cite{kim2017synchrotron}. The details of these steps are described more thoroughly in Appendix~\ref{app:nonideal}. The revised rate of change for the ponderomotive phase is 
\begin{align}
    \frac{d\theta_j}{dz} = 2k_u\eta_j - \frac{k_r}{2}(\vec{x}_j^2k_\beta^2+(\vec{x}_j')^2)
\end{align}
where $k_\beta=1/\bar{\beta}$ is the inverse of the average transverse beta function in the focusing channel, and $\vec{x}_j$ and $\vec{x}_j'$ are the transverse position and angle of the particles. Notably, $\vec{x}_j^2k_\beta^2+(\vec{x}_j')^2$ is a constant of an individual particle's motion. 

\begin{figure*}[!htb]
    \centering
    \includegraphics[width=1.8\columnwidth]{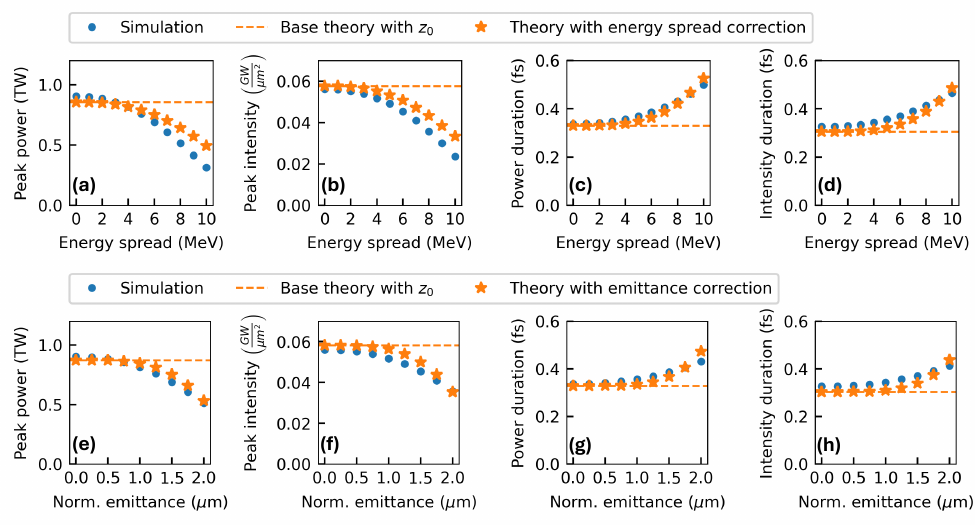}
    \caption{Dependence of field performance metrics on finite energy spread (a-d) and emittance (e-h). }
    \label{fig:espread_emittance_scan}
\end{figure*}

We integrate these equations of motion numerically. We find that, for moderate values of the emittance and energy spread, the bunching profile $b(x)$ continues to be roughly hyperbolic secant in shape, with a modified full-width at half-maximum and peak amplitude. We have numerically determined fitting formulas for the change to these two parameters, 
\begin{align}
    \frac{\text{FWHM}(|b|)_{\sigma_p,\sigma_\epsilon}}{\text{FWHM}(|b|)_0} \simeq 1 + 4\sigma_p^3 + 3\sigma_\epsilon^{4.3}+1.45\sigma_p^{0.78}\sigma_\epsilon^{2.4}\\
    \frac{\text{max}(|b|)_{\sigma_p,\sigma_\epsilon}}{\text{max}(|b|)_0} \simeq 1 - 1.16\sigma_p^{2.6} - 1.13\sigma_\epsilon^{3.8}-3.32\sigma_p^{1.72}\sigma_\epsilon^{4.84}
\end{align}
The scaled energy spread and emittance parameters $\sigma_p$ and $\sigma_\epsilon$, respectively, are given by 
\begin{align}
    \sigma_p &= \frac{\sigma_\eta}{\sqrt{\frac{k_r}{k_u}\kappa\chi\frac{|\ln(1-iz/k_r\sigma_r^2)|}{4\pi}}}\\
    \sigma_\epsilon^2 &= \frac{k_r\epsilon^2}{2k_u\sigma_r^2}\frac{1}{\sqrt{\frac{k_r}{k_u}\kappa\chi\frac{|\ln(1-iz/k_r\sigma_r^2)|}{4\pi}}}
\end{align}
where $\epsilon$ is the geometric emittance assumed to be the same in the $x$ and $y$ planes. The simple fitting formulas reported above are generally valid within 10\% error for $\sqrt{\sigma_p^2+\sigma_\epsilon^2}<0.7$ (see Appendix~\ref{app:nonideal} for more details).

These corrections to the peak and width of the bunching translate into similar corrections for the peak and width of the power and on-axis intensity profiles. In Figure~\ref{fig:espread_emittance_scan} we compare our theoretical model to simulations  from GENESIS 1.3. We scan the normalized emittance and energy spread from 0 to 2 microns and 0 to 10 MeV, respectively, with a fixed 20 $\mu$m beam size. For reference, we note that at the Linac Coherent Light Source, typical values for a 2 kA current beam are 0.4 $\mu$m normalized emittance and 2 MeV slice energy spread. We observe that the simulation results begin to depart from the baseline theoretical predictions around 1 micron normalized emittance and 5 MeV slice energy spread for our particular parameter space. Our theoretical model, corrected for energy spread and emittance, agrees very well with the trends observed in the simulations. It is especially worth noting that these relatively large emittance and energy spread values degrade our field metrics by roughly a factor of 2 at worst, whereas 2 micron normalized emittance and 10 MeV slice energy spread are sufficient to increase the exponential gain length by 20\% and 100\% according to the Ming Xie estimate, respectively \cite{xie1995design}.

\section{Discussion and conclusions}

The 3D scaling laws we have derived for the superradiant mode are instructive for finding the best ways to utilize superradiance. For example, although superradiance has no saturation, the process is limited by the amplification of shot-noise by the linear FEL instability. In fact,  while the superradiant mode propagates over one part of the beam, the rest of the beam undergoes the usual FEL process driven by shot noise, referred to as self-amplified spontaneous emission (SASE). Thus, in practice, the superradiant solution is only useful before SASE saturates in the rest of the beam. Therefore, one should operate the FEL in a way that increases the SASE saturation length without strongly impacting the superradiant mode. Our results indicate that the superradiant solution is a weak function of the beam size, whereas the SASE gain length grows like $\sigma_r^{2/3}$. This implies that by increasing the beam size one can exponentially suppress SASE with minimal impact on superradiance. Similarly, we have found that the superradiant mode is much more tolerant of non-ideal effects like emittance and energy spread than SASE. These results are consistent with the simulation and experimental results of \cite{franz2024}. As another example of the power of the 3D scaling laws, our formulas provide a promising scaling for producing short pulses at high photon energies. For example, the diffraction-dominated superradiant pulse duration has a more favorable scaling with the beam energy and current compared to the SASE cooperation length, proportional to $ \lambda_r\sqrt{\gamma/I}$ for superradiance and $\lambda_r\gamma/I^{1/3}$ for the high-gain regime. These considerations can guide the field in the pursuit of ever shorter, higher power pulses. 

In conclusion, we have presented the physics of a previously unstudied regime of the free-electron laser: diffraction-dominated superradiance. Through 3D simulations and a self-consistent asymptotic non-linear theory, we have shown that after sufficient propagation distance, superradiant free-electron lasers become limited by diffraction. In the limit that $z/k_r\sigma_r^2>1$, the superradiant dynamics slow to a logarithmic rate, with a pulse duration that varies as $1/\sqrt{|\ln(1-iz/k_r\sigma_r^2)|}$ instead of $1/\sqrt{z}$, and a peak power which begins to grow linearly rather than quadratically. These findings explain features of recent experiments with superradiant FELs and provide crucial information towards understanding their ultimate performance limits.

This work was supported by the U.S. Department of Energy, Office of Science, Office of Basic Energy Sciences under
Contract No. DE-AC02-76SF00515, and by the U.S. Department of Energy, Office of Science, Office of Basic Energy Sciences Accelerator and Detector Research Program. R. Robles acknowledges the support of the SLAC Siemann Fellowship. The authors thank Zhirong Huang, Erik Hemsing, Nathan Majernik, Nicholas Sudar, and Filippo Sottocorona for helpful discussions. 

\appendix

\section{1D self-similar solution with diffraction correction}\label{app:scaled_rho}

\begin{figure*}[htb!]
    \centering
    \includegraphics{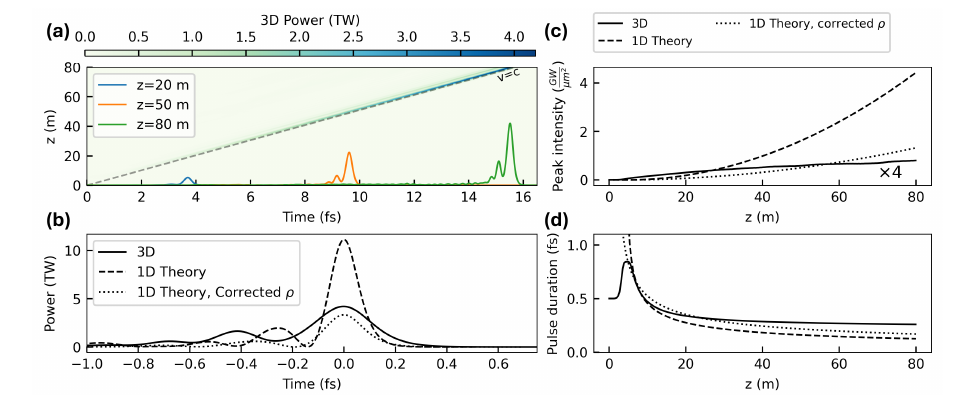}
    \caption{Same structure as Fig.~\ref{fig:1D_3D_comparison}. 3D simulations of an 80 meter undulator are compared to the 1D self-similar solution, now including a version scaled according to the diffraction-corrected value of the Pierce parameter.}
    \label{fig:1D_vs_3D_with_correction}
\end{figure*}

Figure~\ref{fig:1D_vs_3D_with_correction} shows similar results as Figure~\ref{fig:1D_3D_comparison}, only now we have added an extra theory line corresponding to the diffraction-rescaled 1D theory. We note that the theory curves are shifted in $z$ to overlap roughly the first $1/\sqrt{z}$-like decay around 10 meters. The rescaling has the effect of reducing the peak power and intensity while increasing the pulse duration, however it does not change the fundamentally different scalings with $z$ between the 1D and 3D cases. The diffraction-corrected 1D power still grows quadratically and the pulse duration still shrinks as $1/\sqrt{z}$, both in clear contrast to the 3D simulations.

\section{Extended derivations}\label{app:derivations}

\subsection{Calculation of the field for unknown bunching}
In this section we will provide a more detailed derivation of the expressions in the main text. We start from the FEL wave equation given in the main text as 
\begin{align}
    \left(\frac{\partial}{\partial z}+\frac{1}{c}\frac{\partial}{\partial t}-\frac{1}{2ik_r}\nabla_\perp^2\right)E_x(z,t,\mathbf{r})=\kappa b(z,t)u(\mathbf{r})
\end{align}
If we move from $t$ to $\zeta=z-ct$, then $\partial_z+\partial_{ct}=\partial_z$. Furthermore, we'll move to angular space transversely by introducing the Fourier transform $\tilde{E}_x(z,t,\mathbf{k})=\int d\mathbf{r}e^{-i\mathbf{k}\cdot\mathbf{r}}E_x(z,t,\mathbf{r})$. The wave equation becomes 
\begin{align}
    \left(\frac{\partial}{\partial z}+\frac{k^2}{2ik_r}\right)\tilde{E}_x(z,\zeta,\mathbf{k})=\kappa b(\zeta)\tilde{u}(\mathbf{k})
\end{align}
where $k^2=\mathbf{k}\cdot\mathbf{k}$. Since we are assuming that $b(z,t)$ is a function only of $\zeta$ in this first analysis, this equation can be solved directly. That solution is 
\begin{align}
    \tilde{E}_x(z,\zeta,\mathbf{k}) = 2ik_r\kappa b(\zeta)\frac{\tilde{u}(\mathbf{k})}{k^2}\left(1-e^{\frac{ik^2z}{2k_r}}\right)
\end{align}
By inverting the Fourier transform we find the spatial representation of the field, 
\begin{align}
    E_x(z,\zeta,\mathbf{r}) = 2ik_r\kappa b(\zeta)\int\frac{d^2\mathbf{k}}{4\pi^2}e^{i\mathbf{k}\cdot\mathbf{r}}\frac{\tilde{u}(\mathbf{k})}{k^2}\left(1-e^{\frac{ik^2z}{2k_r}}\right)
\end{align}
For ease of later calculations it is convenient to specialize this expression to the case of an axisymmetric beam, $u(\mathbf{r})=u(r)$. For that we can write 
\begin{align}
    E_x(z,\zeta,\mathbf{r}) = 2ik_r\kappa b(\zeta)\int\frac{dk}{2\pi}J_0(kr)\frac{\tilde{u}(k)}{k}\left(1-e^{\frac{ik^2z}{2k_r}}\right)
\end{align}
As we explained in the text, the particle dynamics are affected by the on-axis field in the asymptotic diffraction-dominated limit. For a gaussian beam $u(r)=\frac{1}{2\pi\sigma_r^2}e^{-\frac{r^2}{2\sigma_r^2}}$ and $\tilde{u}(k)=e^{-\frac{k^2\sigma_r^2}{2}}$, and the on-axis field is 
\begin{align}
    E_x(z,\zeta,\mathbf{0}) &= ik_r\kappa b(\zeta)\frac{1}{2\pi}\ln\left(1-\frac{iz}{k_r\sigma_r^2}\right)\\
    &\equiv ik_r\kappa b(\zeta)\frac{1}{2\pi}\left|\ln\left(1-\frac{iz}{k_r\sigma_r^2}\right)\right|e^{-i\phi(z)}
\end{align}
where the phase is defined as $\phi(z)=-\text{arg}\left[\ln\left(1-iz/k_r\sigma_r^2\right)\right]$. Similarly, we can write the on-axis field for a hard-edge cylindrical beam with $u(r)=\frac{1}{\pi R^2}$ for $r<R$. In this case $\tilde{u}(k)=2J_1(kR)/kR$, and 
\begin{widetext}
\begin{align}
    E_x(z,\zeta,\mathbf{0}) &= 2ik_r\kappa b(\zeta)\frac{1}{4\pi}\left[\Gamma_0\left(\frac{ik_rR^2}{2z}\right)-\frac{2iz}{k_rR^2}\left(1-e^{\frac{-ik_rR^2}{2z}}\right)\right]\\
    &\simeq ik_r\kappa b(\zeta)\frac{1}{2\pi}\left|\ln\left(1-\frac{3iz}{k_rR^2}\right)\right|e^{-i\phi(z)}
\end{align}
\end{widetext}
where now $\phi(z)=-\text{arg}\left[\ln\left(1-3iz/k_rR^2\right)\right]$ -- in other words, the on-axis field in the hard-edge case is very similar to that of the gaussian case with the substitution $\sigma_r\rightarrow R/\sqrt{3}$

\subsection{Scaling and solution of the particle motion}

\begin{figure}[h!]
    \centering
    \includegraphics{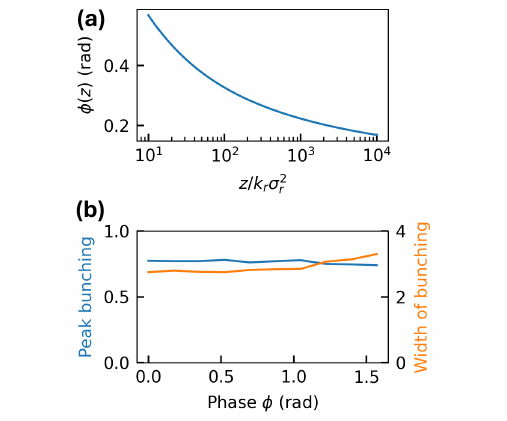}
    \caption{Impact of the field phase $\phi(z)$ on the particle dynamics. (a) $\phi(z)$ for $z/k_r\sigma_r^2$ varying from $10$ to $10^4$. (b) Peak bunching and bunching width from macroparticle tracking simulations for varying phase $\phi(z)$.}
    \label{fig:ignore_the_phase}
\end{figure}

Given this field, we can proceed to studying the particle dynamics to self-consistently determine the bunching. The pendulum equations for the particles are
\begin{align}
    \frac{d\eta_j}{dz} &= -\chi\left(E_xe^{-i\theta_j}+\text{c.c.}\right)\\
    \frac{d\theta_j}{dz} &= 2k_u\eta_j
\end{align}
We are interested in solving the dynamics of a single slice at some fixed ponderomotive phase variable $\theta_0$. The reason we only need to track the dynamics of one slice is precisely because the bunching depends only on $\zeta$ -- it means that when properly scaled, the dynamics of all slices are equivalent. In terms of the ponderomotive phase, we may write $\zeta=z-ct=\frac{\theta_0-k_uz}{k_r}$, implying that for fixed $\theta_0$ we can relate the derivatives $\frac{d}{dz}=-\frac{k_u}{k_r}\frac{d}{d\zeta}$. Thus we can write 
\begin{align}
    \frac{d\eta_j}{d\zeta} &= \frac{k_r^2}{k_u}\kappa\chi \frac{|\ln(1-iz/k_r\sigma_r^2)|}{2\pi}\left(ib(\zeta)e^{-i(\theta_j+\phi(z))}+\text{c.c.}\right)\\
    \frac{d\theta_j}{d\zeta} &= -2k_r\eta_j
\end{align}
The form of these equations suggests scaling to the variables $x=2k_r\zeta\sqrt{\frac{k_r}{k_u}\kappa\chi\frac{|\ln(1-iz/k_r\sigma_r^2)|}{4\pi}}$ and $p_j=\eta_j/\sqrt{\frac{k_r}{k_u}\kappa\chi\frac{|\ln(1-iz/k_r\sigma_r^2)|}{4\pi}}$, where we assume that during the passage of the short superradiant pulse over this slice of the beam the logarithm may be treated as constant. Note that this means that the on-axis field phase $\phi(z)$ is also roughly constant over the timescale of the electron dynamics. These scalings are reminiscent of those used in 1D superradiant theory, only in the 1D case it would have gone like $\sqrt{z}\zeta$ instead of $\sqrt{\ln(z)}\zeta$. In this scaling we have 
\begin{align}
    \frac{dp_j}{dx} &= ib(x)e^{-i(\theta_j+\phi(z))}+\text{c.c.}\label{eqn:scaled_pend_1}\\
    \frac{d\theta_j}{dx} &= -p_j\label{eqn:scaled_pend_2}
\end{align}
These equations can now be used to self-consistently describe the bunching in the beam. Before we do that analytically, we should discuss the phase $\phi(z)$. Just like the amplitude, the phase varies logarithmically with $z$, so it can similarly be treated as constant during the dynamics of one slice. Even more than that, as $z\rightarrow\infty$, this phase approaches zero. We can see the phase and its impact on the particle dynamics in Fig.~\ref{fig:ignore_the_phase}. In panel (a) we plot the phase $\phi(z)=-\text{arg}(\ln(1-iz/k_r\sigma_r^2))$ for $10<\frac{z}{k_r\sigma_r^2}<10^4$. We see that it is slowly but monotonically decreasing, and over this range of three orders of magnitude it varies from 0.55 down to 0.15. To place that magnitude in context we show in panel (b) the results of tracking simulations employing the scaled pendulum equations~\eqref{eqn:scaled_pend_1}-\eqref{eqn:scaled_pend_2}. In particular we initiate 4,096 particles and solve these equations numerically for phase $\phi$ varying from 0 to $\pi/2$ and for a fixed initial bunching of 0.1\%. In panel (b) we are plotting the peak bunching and the width of the rising edge of that bunching in $x$. We observe that in general the dynamics vary weakly with $\phi$, and especially for $\phi<0.5$ radians there is nearly no difference, most notably a 3\% increase in the width of the bunching which is negligible for our current considerations. With these justifications we will proceed simply ignoring the phase term $\phi(z)$.

To study the dynamics of the particles up to the first peak of the superradiant pulse we can employ the nonlinear collective variables method developed in \cite{hemsing2020simple}. To do so we need to introduce the collective energy modulation $P=\langle p_je^{i\theta_j}\rangle$. We can then write equations of motion for the bunching and for $P$,
\begin{align}
    \frac{db}{dx} &= i\left\langle \frac{d\theta_j}{dx}e^{i\theta_j}\right\rangle\\
    &=-iP\\
    \frac{dP}{dx} &= \left\langle \frac{dp_j}{dx}e^{i\theta_j}+ip_j\frac{d\theta_j}{dx}e^{i\theta_j}\right\rangle\\
    &=ib-ib^*\langle e^{2i\theta_j}\rangle-i\langle p_j^2e^{i\theta_j}\rangle
\end{align}
It has been shown by Hemsing \cite{hemsing2020simple} that up through the early saturation regime of the FEL, the following scalings approximately hold: $\langle e^{2i\theta_j}\rangle\simeq \langle e^{i\theta_j}\rangle^2=b^2$ and $\langle p_j^2e^{i\theta_j}\rangle\simeq \langle p_j^2\rangle \langle e^{i\theta_j}\rangle=\langle p_j^2\rangle b$. The energy spread $\langle p_j^2\rangle$ can be determined by evaluating its equation of motion,
\begin{align}
    \frac{d\langle{p_j^2\rangle}}{dx} &= 2\langle ibp_je^{-i\theta_j}-ib^*p_je^{i\theta_j}\rangle\\
    &=2i\left(bP^*-b^*P\right)\\
    &=2\frac{d|b|^2}{dx}
\end{align}
from which it follows the conservation law $\frac{d}{dx}\left(\langle p_j^2\rangle-2|b|^2\right)=0$. Assuming vanishing initial energy spread and bunching, this means that that $\langle p_j^2\rangle =2|b|^2$. Plugging this into the expression for $\frac{dP}{dx}$ and combining it with the expression for $\frac{db}{dx}$, we find a single second-order nonlinear differential equation for the bunching
\begin{align}
    \frac{d^2b}{dx^2}-b+3b|b|^2=0
\end{align}
This equation is very nearly the same nonlinear Duffing equation studied by Hemsing in \cite{hemsing2020simple}. We will approach it in the same way he did. We start by noting that in the linear regime (the leading edge of the superradiant pulse), this equation reads $\frac{d^2b}{dx^2}=b$, with exponential solutions of the form $e^{\pm x}$. Of particular note is that the solution is purely real, or more accurately it has constant phase. In Hemsing's case, his linear regime solution was the typical linear FEL solutions which have linearly varying phase, so he assumed his field to have a linear phase as well. Analogously, we will assume that the phase of $b$ is constant through the first synchrotron oscillation (first superradiant peak), writing $b=Be^{i\phi_0}$. In that case we have 
\begin{align}
    \frac{d^2B}{dx^2}-B+3B^3=0
\end{align}
and this is now precisely the nonlinear Duffing equation studied by Hemsing. The solution is given in terms of the cosine-like Jacobi elliptic function as 
\begin{align}
    B(x) = Y\textbf{cn}\left(\kappa^2x+\psi|n\right)
\end{align}
The parameters $Y$, $\kappa$ and $n$ are defined by 
\begin{align}
    Y^2 = \frac{\alpha\pm\sqrt{(\alpha-\beta B_0^2)^2+2\beta(B_0')^2}}{\beta}
\end{align}
and
\begin{align}
    \kappa^2 = \beta Y^2-\alpha\hspace{2cm} n=\frac{\beta Y^2}{2\kappa^2}
\end{align}
where $\alpha=1$ and $\beta=3$ for our case. $B_0$ and $B_0'$ are the initial conditions, which we will find to have a minimal impact on our results. In particular, we will assume $B_0,B_0'\ll1$ so that we may write approximately
\begin{align}
    Y^2&\simeq \frac{2}{3}+(B_0')^2-B_0^2\\
    \kappa^2&\simeq 1+3((B_0')^2-B_0^2)\\
    n &\simeq 1 -\frac{3}{2}((B_0')^2-B_0^2)
\end{align}
For small initial conditions, then we have $Y\simeq\sqrt{2/3}$, $\kappa\simeq 1$, and $n\simeq 1$. For $n\simeq 1$, the cosine-like Jacobi elliptic function behaves like a hyperbolic secant, $\textbf{cn}(x|n\simeq 1)\simeq \text{sech}(x)$. These allow us to write the approximate result 
\begin{align}
    B(x)\simeq\sqrt{\frac{2}{3}}\text{sech}\left(x\right)
\end{align}
where we have neglected the offset $\psi$ since we are not concerned with the specific location of the peak, only its height and width. 

\subsection{Properties of the field in the zeroth order solution}

It follows that we can write the field up to a phase factor as 
\begin{align}
    E_x(z,\zeta,\mathbf{r}) &= \text{sech}\left(2k_r\zeta\sqrt{\frac{k_r}{k_u}\kappa\chi\frac{|\ln(1-iz/k_r\sigma_r^2)|}{4\pi}}\right)\nonumber\\ 
    &\times2ik_r\kappa\sqrt{\frac{2}{3}}\int\frac{dk}{2\pi}J_0(kr)\frac{\tilde{u}(k)}{k}\left(1-e^{\frac{ik^2z}{2k_r}}\right)
\end{align}
Let us now compute several properties of this field. First, both the on-axis intensity and the power will depend on $\zeta$ as the square of the sech function. The full-width at half-maximum of $\text{sech}^2(x)$ is $2\text{cosh}^{-1}(\sqrt{2})$, so it follows that the FWHM temporal width is 
\begin{align}
    \text{FWHM} = \frac{2\text{cosh}^{-1}(\sqrt{2})}{\omega_r}\sqrt{\frac{\gamma_0}{|\ln(1-iz/k_r\sigma_r^2)|}\frac{I_A}{I}\frac{1+\frac{K^2}{2}}{K^2\text{[JJ]}^2}}
\end{align}
Note that this is the same for the hard-edge cylindrical beam as long as $\sigma_r\rightarrow R/\sqrt{3}$, as before. 

The maximal on-axis intensity on the pulse is given by 
\begin{align}
    I_\text{max} &= 2\epsilon_0c|E_x(z,0,\mathbf{0})|^2\\
    &=\frac{1}{3\epsilon_0c}\left[\frac{IK\text{[JJ]}k_r}{4\pi\gamma_0}\left|\ln\left(1-\frac{iz}{k_r\sigma_r^2}\right)\right|\right]^2
\end{align}
and the peak power is 
\begin{align}
    P_\text{max} &= \int 2\epsilon_0c|E_x(z,0,\mathbf{r})|^2d\mathbf{r}\\
    &=\frac{4}{3\epsilon_0c}\left(\frac{IK\text{[JJ]}k_r}{\gamma_0}\right)^2\int\frac{dk}{2\pi}\frac{\tilde{u}(k)^2}{k^3}\sin^2\left(\frac{k^2z}{4k_r}\right)
\end{align}
For the gaussian beam this is 
\begin{widetext}
\begin{align}
    P_\text{max} &= \int 2\epsilon_0c|E_x(z,0,\mathbf{r})|^2d\mathbf{r}\\
    &=\frac{4\pi\sigma_r^2}{3\epsilon_0c}\left(\frac{IK\text{[JJ]}k_r}{4\pi\gamma_0}\right)^2\left[\frac{z}{k_r\sigma_r^2}\arctan\left(\frac{z}{2k_r\sigma_r^2}\right)-2\text{coth}^{-1}\left(1+\frac{8k_r^2\sigma_r^4}{z^2}\right)\right]\\
    &\simeq\frac{4\pi\sigma_r^2}{3\epsilon_0c}\left(\frac{IK\text{[JJ]}k_r}{4\pi\gamma_0}\right)^2\left[\frac{z}{k_r\sigma_r^2}\frac{\pi}{2}-2\ln\left(\frac{z}{k_r\sigma_r^2}\frac{e}{2}\right)\right]
\end{align}
\end{widetext}
For a hard-edge beam the integral can be evaluated but the result is a long sequence of hypergeometric functions. We will just give the asymptotic approximation:
\begin{align}
    P_\text{max} &= \frac{2\pi R^2}{3\epsilon_0c}\left(\frac{IK\text{[JJ]}k_r}{4\pi\gamma_0}\right)^2\left[\pi\frac{z}{k_rR^2}-\ln\left(2e^{\gamma-\frac{5}{3}}\frac{z}{k_rR^2}\right)\right]
\end{align}
where $\gamma$ is Euler's constant.

\section{Dependence of superradiant source point on beam size}\label{app:z0_vs_beamsize}

\begin{figure}
    \centering
    \includegraphics[width=0.8\columnwidth]{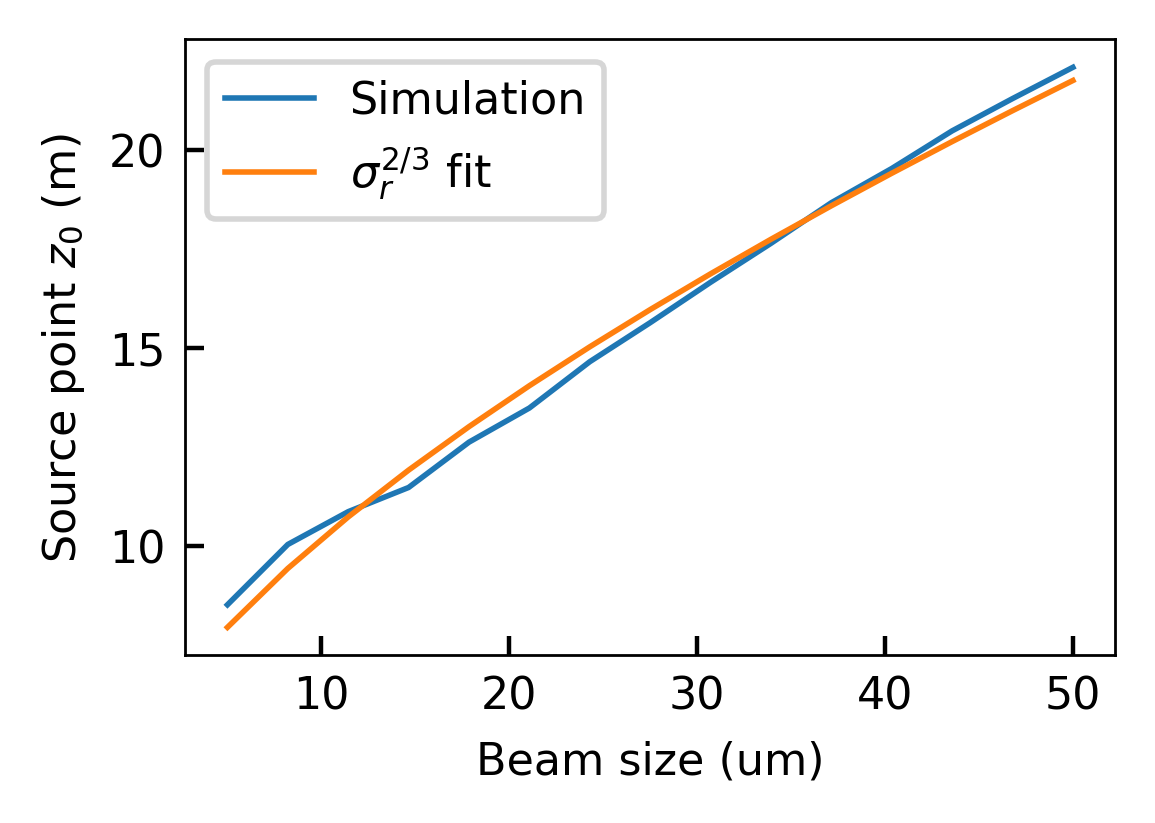}
    \caption{Dependence of $z_0$ in the simulation on the beam size, alongside a $\sigma_r^{2/3}$ fit.}
    \label{fig:z0_vs_beamsize}
\end{figure}

Figure~\ref{fig:z0_vs_beamsize} shows the dependence of the source point $z_0$ on the beam size for the simulations from Figure~\ref{fig:vs_beamsize}. As argued in the main text, $z_0$ is expected to correspond approximately to the saturation length driven by the seed pulse. As a result we expect it to scale inversely with the high gain Pierce parameter $\rho$, which is itself proportional to $\sigma_r^{-2/3}$. Thus alongside the empirically derived $z_0$ values we also plot a fit to $\sigma_r^{2/3}$, which we see captures the behavior of the curve well.

\section{Impact of emittance and energy spread}\label{app:nonideal}

For the purpose of using these formulas in practical experimental settings it is important to understand the impact of non-ideal beam effects like emittance and energy spread. Accounting for the energy spread is in principle as simple as solving the scaled pendulum equations 
\begin{align}
    \frac{dp_j}{dx} &= ib(x)e^{-i\theta_j}+\text{c.c.}\\
    \frac{d\theta_j}{dx} &= -p_j
\end{align}
only now allowing the coordinates $p_j$ to have a non-zero spread. While it is unclear how to do this analytically, it is simple enough to solve these equations numerically and determine fitting functions to the amplitude and full-width at half-maximum of $b(x)$ as a function of a gaussian-distributed energy spread $\sigma_p$. This approach assumes that the basic hyperbolic secant form of the bunching doesn't change as energy spread is introduced, which should be studied in more depth generally but seems to be true for relatively small values.

Accounting for emittance requires a modification to the pendulum equations as we've written them so far. The primary effect of emittance is to further slow the longitudinal velocity of the electrons due to the particles' thermal transverse velocities. We will consider the case that the beam has finite emittance but is confined to a constant size by artificially enhanced natural undulator focusing, which gives rise to the same equations of motion as the case of strong focusing by a FODO lattice \cite{kim2017synchrotron}. In such a focusing channel, the transverse coordinates undergo betatron oscillations of the form
\begin{align}
    x_\beta(z) &= x_\beta(0)\cos(k_\beta z) + \frac{x_\beta'(0)}{k_\beta}\sin(k_\beta z)\\
    y_\beta(z) &= y_\beta(0)\cos(k_\beta z) + \frac{y_\beta'(0)}{k_\beta}\sin(k_\beta z)
\end{align}
where $k_\beta=1/\bar{\beta}$, where $\bar{\beta}$ is the average beta function of the beam in the focusing channel. It was shown in \cite{kim2017synchrotron} that the key change to the FEL equations from this betatron motion is to change the rate of change of the ponderomotive phase in the following way:
\begin{align}
    \frac{d\theta_j}{dz} = 2k_u\eta_j - \frac{k_r}{2}(\vec{x}_j^2k_\beta^2+(\vec{x}_j')^2)
\end{align}
Notably, the combination $\vec{x}_j^2k_\beta^2+(\vec{x}_j')^2$ is a constant of motion in a smooth focusing channel which can be written as $2k_\beta \mathcal{J}_j$ as in \cite{kim2017synchrotron}, where $\mathcal{J}_j$ is the conserved single particle action. The ensemble average of that action over the beam is equal to the geometric emittance of the beam, $\langle \mathcal{J}_j\rangle=\epsilon$, which we assume to be the same in both planes.

Introducing the scalings as before, we can write the pendulum equations now as 
\begin{align}
    \frac{dp_j}{dx} &= ib(x)e^{-i\theta_j}+\text{c.c.}\label{eqn:nonideal_pend_p}\\
    \frac{d\theta_j}{dx} &= -p_j + \frac{\sigma_\epsilon^2}{2} (\vec{\chi}_j^2+(\vec{\chi}_j')^2)\label{eqn:nonideal_pend_th}
\end{align}
where
\begin{align}
    \sigma_\epsilon^2 = \frac{k_r\epsilon^2}{2k_u\sigma_r^2}\frac{1}{\sqrt{\frac{k_r}{k_u}\kappa\chi\frac{|\ln(1-iz/k_r\sigma_r^2)|}{4\pi}}}
\end{align}
We have introduced $\vec{\chi}_j\equiv\vec{x}_j/\sigma_r$ and $\vec{\chi}_j'\equiv \vec{\chi}_j'/\sigma_r'=\vec{\chi}_j'/(\epsilon/\sigma_r)$, meaning that these are now (for a gaussian beam distribution) random variables with standard deviation of 1. 

\begin{figure}[h!]
    \centering
    \includegraphics[width=0.9\columnwidth]{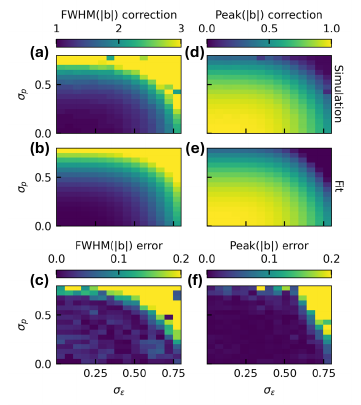}
    \caption{Dependence of numerically tracked FWHM and peak of the bunching as a function of energy spread and emittance. (a-c) deal with the FWHM, (d-f) with the peak value. (a) and (d) show simulation results, (b) and (e) show the result obtained from the fitting formula, and (c) and (f) show the relative error.}
    \label{fig:fit_figure}
\end{figure}

We have solved equations~\eqref{eqn:nonideal_pend_p} and \eqref{eqn:nonideal_pend_th} numerically for $\sigma_\epsilon$ and $\sigma_p$ ranging from 0 to 0.8 each. In general the bunching continues to behave like a hyperbolic secant through its first peak, with the effect of either energy spread or emittance being to increase the duration of the profile while decreasing the amplitude. In particular we have determined the following fitting formulas for the amplitude and width of the bunching: 
\begin{align}
    \frac{\text{FWHM}(|b|)_{\sigma_p,\sigma_\epsilon}}{\text{FWHM}(|b|)_0} \simeq 1 + 4\sigma_p^3 + 3\sigma_\epsilon^{4.3}+1.45\sigma_p^{0.78}\sigma_\epsilon^{2.4}\\
    \frac{\text{max}(|b|)_{\sigma_p,\sigma_\epsilon}}{\text{max}(|b|)_0} \simeq 1 - 1.16\sigma_p^{2.6} - 1.13\sigma_\epsilon^{3.8}-3.32\sigma_p^{1.72}\sigma_\epsilon^{4.84}
\end{align}
These simple results are generally valid within 10\% error for $\sqrt{\sigma_p^2+\sigma_\epsilon^2}<0.7$, as can be seen in Figure~\ref{fig:fit_figure}.

\bibliography{refs}

\end{document}